# Tuning the coherent propagation of organic exciton-polaritons through dark state delocalization

[1,2]Raj Pandya, [1]Arjun Ashoka, [3,4]Kyriacos Georgiou, [1]Jooyoung Sung, [3]Rahul Jayaprakash, [5]Scott Renken, [6,7]Lizhi Gai, [7]Zhen Shen, [1]Akshay Rao and [5]Andrew Musser*

[1]Cavendish Laboratory, University of Cambridge, J.J. Thomson Avenue, CB3 0HE, Cambridge, United Kingdom

[2]Laboratoire Kastler Brossel, École Normale Supériéure-Université PSL, CNRS, Sorbonne Université, College de France, Paris 75005, France

[3] Department of Physics and Astronomy, University of Sheffield, Sheffield S3 7RH, United Kingdom

[4] Department of Physics, University of Cyprus, P.O. Box 20537, Nicosia, 1678, Cyprus

[5] Department of Chemistry and Chemical Biology, Cornell University, 14853 Ithaca, NY, USA

[6] Key Laboratory of Organosilicon Chemistry and Material Technology, Ministry of Education, Hangzhou Normal University, Hangzhou, 311121, China

[7] State Key Laboratory of Coordination and Chemistry, School of Chemistry and Chemical Engineering, Nanjing University, Nanjing, 210046, China

correspondence: ajm557@cornell.edu

**Abstract**

While there have been numerous reports of long-range polariton transport at room-temperature in organic cavities, the spatio-temporal evolution of the propagation is scarcely reported, particularly in the initial coherent sub-ps regime, where photon and exciton wavefunctions are inextricably mixed. Hence the detailed process of coherent organic exciton-polariton transport and in particular the role of dark states has remained poorly understood. Here, we use femtosecond transient absorption microscopy to directly image coherent polariton motion in microcavities of varying quality factor. We find the transport to be well-described by a model of band-like propagation of an initially Gaussian distribution of exciton-polaritons in real space. The velocity of the polaritons reaches values of ~$0.65 \times 10^6$ m s$^{-1}$, substantially lower than expected from the polariton dispersion. Further, we find that the velocity is proportional to the quality factor of the microcavity. We suggest this unexpected link between the quality-factor and polariton velocity and slow coherent transport to be a result of varying admixing between delocalised dark and polariton states.

*Keywords: exciton-polariton, coherent transport, energy transport, dark states, Q-factor, TA microscopy*

Introduction

The interaction of light with matter (atoms, excitons, vibrational dipoles, *etc*) trapped inside an optical resonator is governed by three key rates: the rate at which light and matter transfer energy ($g$), the rate at which light escapes the cavity ($\kappa$) and the rate at which the matter itself loses its polarisation (exciton-photon dephasing) ($\zeta$). Strong coupling occurs when the rate of energy transfer, $g$, is considerably larger than $\kappa$ and $\zeta$ ($g \gg \kappa, \zeta$).[1–3] Light and matter then exchange energy periodically (with period $\frac{2}{g\pi}$) until energy escapes the system – usually by the emission of photons from the cavity. This rapid exchange results in the formation of a new light-matter hybrid state, the polariton. Where the light-matter interaction involves the electronic absorption of a semiconductor exciton, the states formed are specifically exciton-polaritons. Though chiefly studied with inorganic semiconductors,[4] exciton-polaritons can also be achieved with molecular materials.[5,6] Their large oscillator strength can yield extreme Rabi splittings – a measure of the interaction strength – from 100's of meV to >1 eV.[7–11]

In the last decade, these organic systems have been subject to increasing interest for the remarkable effects found to accompany polariton formation. In addition to exhibiting Bose-Einstein condensation, lasing and superfluidity,[12–15] organic polaritons have been reported to result in significant changes to bulk material properties from work function to energy and charge carrier transport,[16–19] and alterations to molecular photophysics from isomerization to spin interconversion.[20–22] These advantages persist in devices, where they permit efficient emission or enhanced light-harvesting.[8,23–25] One of the most promising polaritonic properties that may be utilised in optoelectronic devices is their large energy transfer or propagation length, which ranges from 100's of nm to 10's of μm depending on cavity architecture.[17,26–33] Based on their reduced mass compared to excitons, polaritons are predicted to reach group-velocities of $10^7$ m s$^{-1}$, with their in-plane diffusion lengths chiefly limited by their short lifetimes.

While long-range transport indirectly validates this prediction,[30,31] nearly all studies have been performed in the steady state with little understanding of the transport dynamics. A critical question is whether the reported propagation is purely polaritonic or mediated by intracavity dark states. Within the standard Tavis-Cummings model, for N dipoles that couple to the photonic mode, there will result 2 bright states—the polaritons—and N-1 dark states. In bulk organic microcavities, the resulting dark states are expected to dominate the population by a factor ~$10^5$.[34–36] In addition to a vastly greater density of states (DoS), these dark states (also termed the 'exciton reservoir') boast a significantly longer lifetime due to their purely excitonic rather than partially photonic character. Accordingly, they are likely to control most dynamical processes within a microcavity. Thus most reports of organic polariton emission typically follow the population dynamics and profile of the dark states and reflect their various scattering pathways into bright polaritons.[12,37–39] Moreover, these dynamics typically also closely agree with the dynamics of uncoupled excitons in bare organic films, suggesting that the rate of population transfer from reservoir states to the bright polaritons should be slow relative to intrinsic exciton decay channels.[22,37,40]

This interplay between bright polaritons and dark states essentially underpins microcavity functional properties, from transport to chemistry,[41–44] especially when measured in the steady state or on long timescales. The predominance of dark states makes it extremely challenging to identify the unique effects/properties of polaritons themselves. The best prospect for doing so is to measure their properties within the 'coherent' regime in which the exciton-photon superposition state has not yet had sufficient time to dephase into localized molecular states. This has only been achieved once using ultrafast (<30 fs) transient absorption spectroscopy,[45] with subsequent studies focused on low-quality-factor metallic cavities and picosecond dynamics,[46–48] where any spectral signatures should reflect dark states and related optical artefacts.[49–51] Similarly, the one previous time-resolved microscopic study of organic polariton transport examined metallic J-aggregate microcavities in the 1-10 ps regime,[32] which standard models of polaritons maintain is controlled by dark states. Intriguingly, the transport velocity reported was markedly higher than would be expected for pure excitonic diffusion (reaching $0.2 \times 10^6$ m s$^{-1}$), but significantly below the expected group velocity. It remains unclear what role nominally stationary and localized dark states play in this process, or how the transport velocity can be tuned or enhanced.

Here, we overcome these challenges using femtosecond transient absorption spectroscopy and microscopy to study strongly coupled BODIPY dye molecules in dielectric mirror-based cavities over a range of $Q$-factors well above those achieved in all-metal cavities. By employing a technique with ~10 fs time resolution and ~10 nm spatial precision we directly visualise polariton propagation and dynamics in the sub-ps coherent regime. We show that even on these timescales, polariton transport remains slower than the maximum group velocity, but can be described by a simple model of coherent propagation of an initially Gaussian distribution. Unexpectedly, we find a direct relationship between the microcavity $Q$-factor, (which is proportional to the photon lifetime), and the polariton velocity. We propose that this effect can be rationalised by considering the role of disorder within these systems. In typical molecular semiconductors, molecular disorder inhomogeneously broadens the ground- and excited-state energy distribution to well beyond the linewidths of typical Fabry-Perot cavity modes. We posit that when such an inhomogeneously broadened system is placed into the strong-coupling regime, not all the molecules in the ensemble strongly couple, and those most likely to do so are most resonant with the photonic mode. Effectively, then, the relatively narrow photonic mode selects out a narrower population of absorbers. Critically, the selected states do not possess identical energies, but rather exhibit a distribution that approximates the photonic linewidth. In molecular systems with significant intrinsic disorder, this means that the cavity *Q*-factor is a powerful yet unexplored tool to tune the energetic distribution of the N-1 dark states that arise from strong coupling. Consistent with recent theoretical work and common findings in organic semiconductors, we find that as the energetic disorder within this dark-state manifold decreases, the states become more highly delocalized and coherent. This effect permits an unexpectedly rapid channel of population exchange with the rapidly diffusing bright polariton states, resulting in long-range, *Q*-factor-controlled transport. Not only do these results demonstrate a direct means to tune the polariton transport velocity, they also provide a unique platform to study and control the dark states, which may play an even greater role in polariton photodynamics than previously considered.

## **Results**

We employ a series of dielectric Fabry-Perot microcavities containing BODIPY dye molecules as a model exciton-polariton system. Owing to their photostability, large oscillator strength and chemical tunability, BODIPY dyes have proved an optimal platform to explore fundamental polariton physics from dark state scattering and donor-acceptor energy transfer to condensation.[28,37,52] Here, we used a red-emitting derivative BODIPY-R, dispersed in polystyrene matrix at 10 wt% (Figure 1a). At these concentrations some contribution from excimers and aggregated molecules to the total polariton emission can be expected, but such effects occur on relatively long timescales.[37,53] On the sub-ps timescales treated here, the molecules can be considered fully isolated within the host matrix. The microcavities were prepared using a standard combination of spin-coating and e-beam evaporation techniques (see Methods). The microcavity mirrors consisted alternating layers of 71 nm $Nb_2O_5$ and 107 nm $SiO_2$, for a stopband centred at 635 nm—near the first excitonic resonance of the BODIPY molecules (Figure 1a,b). The organic layer thickness was tuned to ~215 nm to achieve ~55 meV detuning of the cavity mode relative to the excitonic resonance. For time-resolved measurements, the full structures were deposited on 100 µm-thick glass slides and encapsulated in inert environment to minimize optical artefacts and prevent photodegradation.

To provide a reference for the polariton energetic structure, we used the same methodology to prepare cavities on smooth quartz substrates, using $TiO_2/SiO_2$ mirrors and the same organic solution. Figure 1c shows the typical angle-dependent reflectivity of a microcavity with the number of layer pairs within each mirror n = 5.5, revealing clear upper and lower polariton branches. At this concentration, the second vibronic peak of BODIPY-R absorbs too weakly to achieve strong coupling to the cavity mode. The slight perturbation of the dispersion evident at 570 nm is indicative of intermediate coupling, and we consider the entire upper band to represent the upper polariton. We can closely match the full observed dispersion using transfer matrix simulations (symbols) benchmarked against the optical spectra of the microcavity components. This match demonstrates that our optical description of the DBRs and organic active layer is highly accurate.

Based of this description, we fabricated a series of microcavities for transient measurements with n varied between 3.5-6.5, affording a set of structures with progressively narrower transmission spectra and accordingly higher *Q*-factors (Figure 1d). Transfer matrix simulations benchmarked against the mirror reflectivity spectra yield idealized *Q*-factors of 66 to 479, which correspond to bare photon lifetimes of 23 to 165 fs (see **SI, S1** for optical modelling). The same BODIPY-R concentration and spin-coating conditions were used to fabricate every cavity, regardless of *Q*-factor. Our simulations (**SI, S1**) confirm that using the same active layer, we thus attain comparable Rabi splitting and detuning across the entire series.

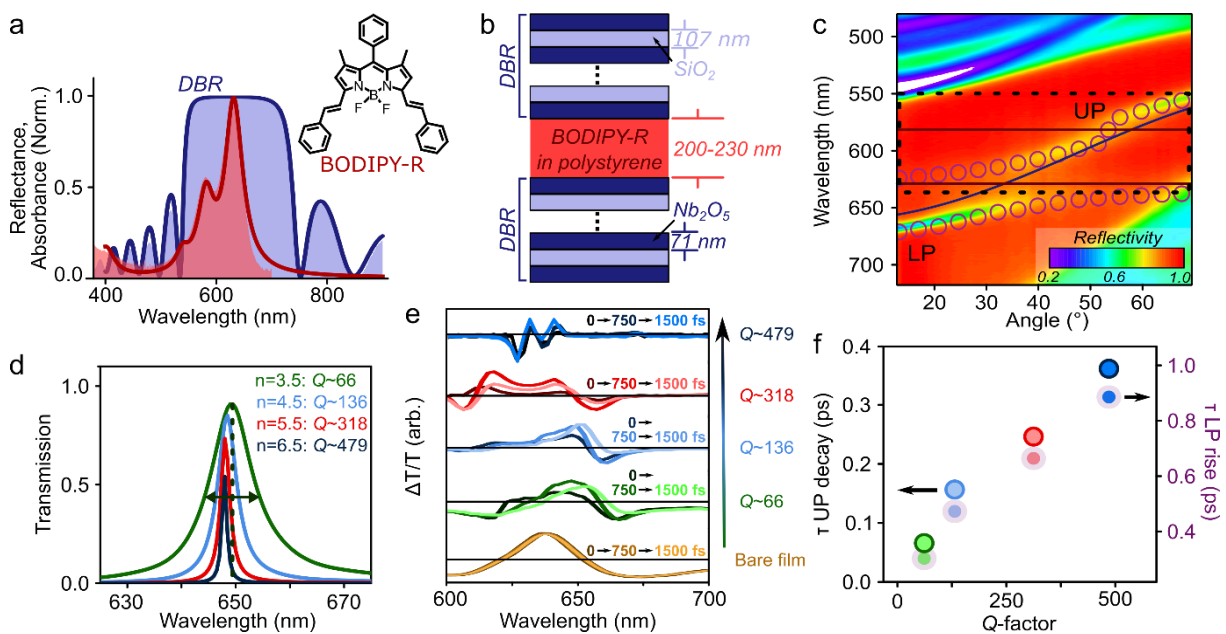

**Figure 1: Optical characterisation of BODIPY-R microcavities. a.** Absorption spectrum of a bare film of BODIPY-R molecules in polystyrene matrix (red) and reflectivity of a DBR with n=7.5 (blue), with superimposed fits (lines) used for transfer matrix simulations. **b.** Schematic of DBR microcavity, consisting of pairs of $Nb_2O_5$/$SiO_2$ alternating layers. BODIPY-R molecules in a polystyrene matrix are sandwiched in the centre of the cavity. **c.** Angle-resolved reflectivity of n = 5.5 microcavity with lower (LP) and upper (UP) polariton branches marked. Circles denote LP and UP dispersions obtained from transfer matrix simulation. The corresponding uncoupled cavity mode and exciton positions are indicated as solid lines. Dashed box shows range of wavelengths covered by pump pulse in fs-TAM experiment and range of wavevectors pumped by the high-NA objective (total range of 128.4°). **d.** Simulated transmission spectra of model (empty) cavities showing the narrowing of spectra with increasing *Q*-factor. **e.** Transient absorption of BODIPY-R dispersed in polystyrene film and microcavities with *Q*-factors ranging from 66 to 479 (n = 3.5 to n = 6.5), excited with a ~10 fs broadband pump pulse (FWHM ~60 nm) centred at 550 nm. In the film the positive feature at 640 nm corresponds to the bleach. In the microcavities, the positive features at 620 to 660 nm match the position of the UP and LP at ~0°. **f.** Decay of UP GSB (unshaded) and rise times of LP GSB (purple shading) extracted from exponential fit of the transient absorption kinetics.

***Identification of photoexcited polariton bands.*** In order to initially characterise the electronic dynamics of the BODIPY-R film and microcavities, broadband transient absorption measurements were performed with a 10 fs pump-pulse. The spectra of the BODIPY-R film, as shown in Figure 1e, are dominated by positive features tracking the ground-state bleach (~640 nm) and stimulated emission (~640nm and ~700 nm) of the photoexcited singlet state, with characteristically weak excited-state absorption. These features decay with an average decay time of 105 ± 20 ps, a typical timescale for excimer formation and migration to excimer sites in BODIPY films at this concentration.[53] The transient spectra of the microcavities are markedly more complex, exhibiting numerous derivative-like features. Similar spectra have been previously reported in other strong-coupled molecular systems and have variously attributed to polaritonic excited-state absorption, polariton contraction from ground-state bleaching, indirect modification of the cavity optical response due to population of dark 'reservoir' states, and a host of non-specific thermal and optical artefacts.[26,27,54,32,45–51] Specific peak assignments are controversial due to the many likely overlapping contributions from multiple origins, as we discuss elsewhere,[51] and are beyond the scope of this report.

Here, our principal concern is to identify spectral bands that uniquely signify the photoexcited polaritons. We exploit the fact that, of all proposed contributing spectral species,[49–51] only those directly related to the polariton population should change with the microcavity *Q*-factor, through the well-known dependence of polariton lifetime $\tau_{pol}^{-1} = A\tau_p^{-1} + B\tau_x^{-1}$, where $\tau_p$ and $\tau_x$ are the lifetimes of the unmixed photonic and excitonic states and *A* and *B* their respective fractions in the wavefunction. Comparing equivalent features for cavities with n = 3.5 – 6.5, we find that the bleach signal (ΔT/T>0) decay at ~620 nm and rise at ~640 nm are strongly *Q*-factor dependent, revealing their photonic character. The photoinduced absorption feature (ΔT/T<0) that is slightly red-shifted from the lowest-energy bleach peak follows the same *Q*-dependence and is a further marker of the polaritons (**SI, S2**). The physical processes that contribute to these dynamics are complex, including the aforementioned artefacts (*Q*-independent), scattering to and from dark states (typically expected to be *Q*-independent), and polariton relaxation (*Q*-dependent), all of which could contribute to the growth/decay of an apparent bleaching band or the decay/growth of an overlapping photoinduced absorption. Regardless of this complexity, the clear *Q*-factor dependence in this regime demonstrates that the signal tracks a partially photonic species and must be diagnostic of the polaritons. The transient signal persists on much longer timescales, including in these spectral regions, but there is no further *Q*-factor dependence beyond the ps regime (see **SI, S3**). This behaviour is indicative of long-lived dark states and non-polaritonic optical effects discussed elsewhere and which dominate the cavity dynamics following initial polariton decay within the first ~500 fs[49–51].

***Real-time tracking of coherent polariton propagation.*** We exploit these spectral signatures and the dynamical insights of our *Q*-factor dependence to interrogate the nature of polariton transport in the initial coherent regime using transient absorption microscopy.[55–57] In this technique, a pump pulse focussed to the diffraction limit (σ~140 nm on SiO$_2$) locally excites a region of the sample. A wide-field counter-propagating probe pulse then reads out the spatial pump-probe signal generated, as a function of time. The spatial size of the signal indicates the spatial extent of the electronic population being probed. By subtracting the spatial pump-probe profile width at a time *t* from that when there is no time delay between the pump and probe ($t_0$) the magnitude of signal propagation can be determined. Though the imaging resolution is limited to the diffraction limit ~300 nm (convolution of pump and probe FWHM), the differential method allows for transport distances to be determined with ~10 nm of spatial *precision*. This figure arises from our ability to resolve even subtly different Gaussian spatial pump-probe signals, thanks to the high signal-to-noise of our experiment and calibration in previous studies[55,57,58]. In order to achieve ultrafast 2D spatial imaging we must sacrifice spectral resolution, so for each sample we characterise the band at 640 nm integrated over a 10 nm window which was established above as a signature of the polariton population. This technique enables direct imaging of the lateral propagation of the polariton population,[32] a phenomenon much less studied than polariton-mediated energy transfer between molecules separated along the cavity thickness.[17,26–29]

We use temporally compressed pump and probe pulses (8 fs and 11 fs, respectively) which give an overall resolution to the experiment of ~15 fs and allow us to capture the crucial initial coherent regime. As a result our pump pulse is spectrally broad (~70 nm FWHM, centred at 560 nm and cut at 630 nm with a razor edge short-pass filter to ensure no leakage into the probe channel). Moreover, the high numerical aperture of our objective (1.4) covers an angular range of ±64.2°, centred on normal incidence. This configuration means that we coherently excite both UP and LP branches over a wide range in our experiment, as well as potentially intracavity exciton states, as indicated by the box in Figure 1c. It is essential to ensure a similar range of states are excited across our microcavity series, as the polariton group velocity and expected transport behaviour depend on the detuning and LP curvature. Great care must be taken because of the local thickness variations between and within such solution-processed films.[51,59] Here, we exploited the spectral position of the derivative-like transient absorption bands ~640 nm as a sensitive marker of cavity thickness.[51] Using our instrument as a narrowband transient absorption spectrometer, we first scanned the sample to identify locations with the same spectral behaviour as in Figure 1e. Based on our previous modelling of such features,[51] we estimate the appearance of this band is only consistent with an active layer thickness of 200-215 nm. This range sharply constrains the possible polariton energetics.

Figure 2a shows typical transient absorption microscopy images for n = 3.5 and n = 6.5 microcavities at selected time delays after photoexcitation. The dotted lines denote the FWHM used to define the signal spatial extent. We highlight three principal observations from these images. 1) There is an increase in the spatial probe signal following photoexcitation at from 0 fs to 125 fs; 2) the increase in the spatial pump-probe signal is larger for the n = 6.5 microcavity than for the n = 3.5 microcavity; and 3) the signal shrinks back to its original dimensions for both cavities from 125 fs to 800 fs. These dynamics, and the underlying fs-TA dynamics, are independent of excitation laser fluence from 40 to 200 µJ cm$^{-2}$ (**SI, S3**). The lifetime of bare BODIPY-R films in these measurements decreases over this range. Equivalent fs-TAM measurements on half-cavities and on thin films of BODIPY-R do not show any measurable population transport on these timescales, confirming that the observed behaviour is a unique consequence of the presence of polaritons (**SI, S4 – S5**).

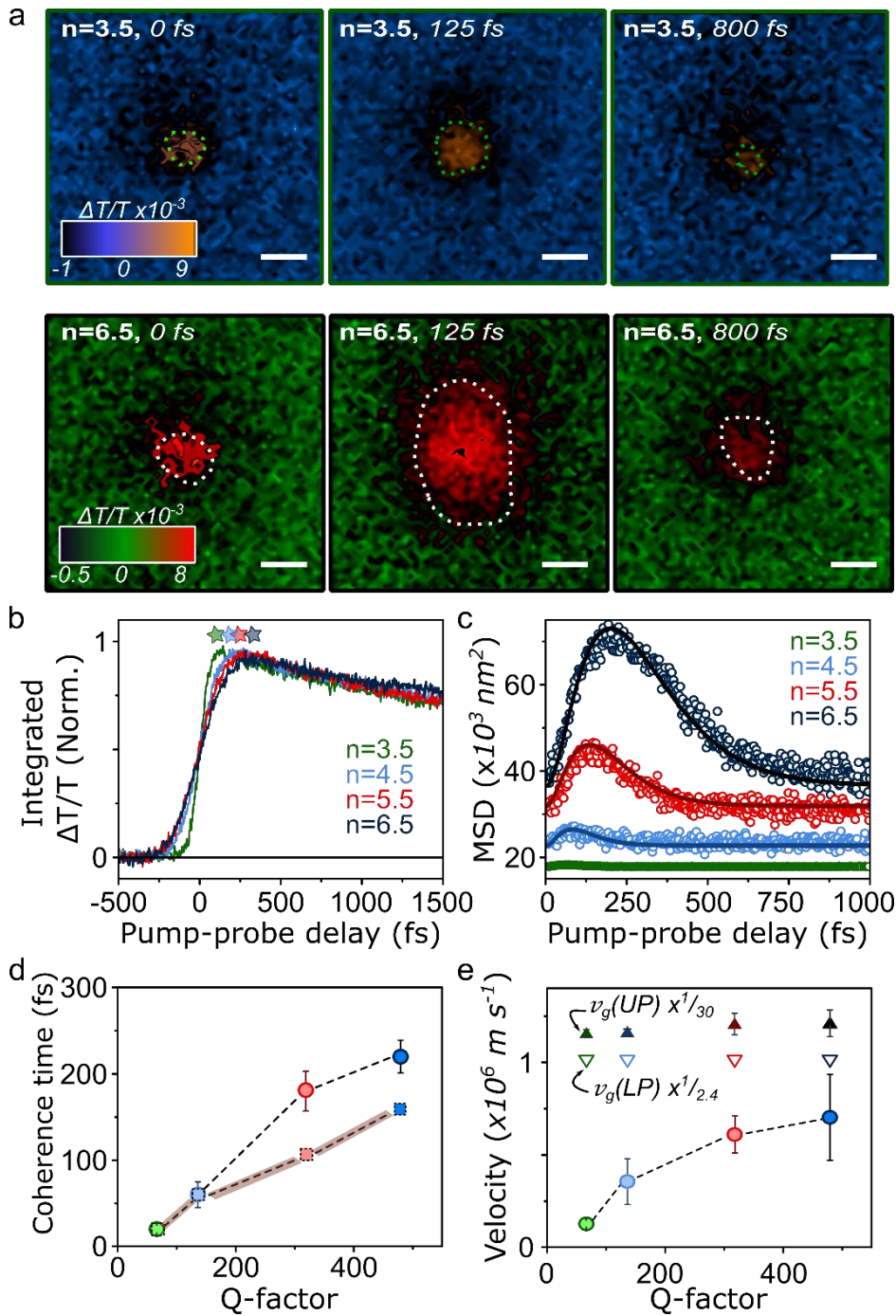

**Figure 2: Transient absorption microscopy images of BODIPY-R molecules in microcavity for n = 3.5 (top) and n = 6.5 (bottom) mirror pairs. a.** Transient absorption microscopy images showing initial expansion up to 125 fs and subsequent contraction. The dotted line indicates the radial Gaussian standard deviation. The scale bar in the images is 500 nm. **b.** Kinetic extracted from centre of spatial pump probe signal for the full cavity series. Stars indicate the signal maximum used to calculate $t_0$ for

each cavity. **c.** Mean-square displacement of signal as a function of time. Circles are raw data and solid line shows fit to Equation 5. **d.** Coherence loss time (circles) extracted from Equation 5 as a function *Q*-factor. Error bars are derived from a minimum of 5 repeat experiments. Squares indicate the corresponding photon lifetime for the cavity. **e.** Exciton-polariton velocity (circles) extracted from fitting Equation 5 to MSD, as a function of *Q*-factor. Experimental error bars are derived from a minimum of 5 repeat experiments at different sample locations. Triangles show expected group velocity of UP (filled) and LP states (open), calculated as described in the main text and scaled for ease of comparison. Error bars reflect standard deviation for cavity thicknesses 200-215 nm.

*Mean-square displacement of polariton population.* For more quantitative analysis we first retrieve the electronic kinetics from the entire spatial pump-probe signal (entire detection area of the probe) as shown in Figure 2b. Within this time window these are in reasonable agreement with the decays obtained from our fs-TA experiments (see **SI, S3**). A salient feature is that as the number of mirror pairs increases so does the signal rise time, from ~20 fs for n = 3.5 to 100 fs for n = 6.5. In order to retrieve the spatial dynamics, the standard deviation for all the transient absorption images is extracted using a radial fitting method (see **SI, S5**). From this the mean-square-displacement (MSD) can be computed, which is a measure of the change in the spatial extent of the exciton-polariton population as a function of time:

$$\text{MSD} = \sigma(t)^2 - \sigma(t_0)^2. \quad \text{(Equation 1)}$$

It is crucial to define $t_0$ for the MSD calculation carefully; if $t_0$ is defined too early then the signal expansion will be artificially enhanced; if it is too late then any ultrafast propagation will be missed. Here, we use the maximum of the LP population as retrieved from the kinetic in Figure 2b to extract $t_0$ and $\sigma(t_0)$. At $t_0$ the MSD reflects the size of the pump pulse. This parameter systematically increases with mirror thickness from 135 nm (n = 3.5) to 190 nm (n = 6.5) (see **SI, S5**). This is due to a refractive index mismatch between the mirrors and the immersion medium for the high N.A. objective which is designed for microscope cover glass (refractive index ~1.52). As we increase the number of mirror pairs, the average refractive index of the cavity increases and deviates further from the design value, introducing systematic variation in the spot size at $t_0$. To eliminate any fitting bias, we independently characterise this value by scanning the pump spot across fluorescent beads on half cavities of varying numbers of mirror pairs. Agreement with these values confirms that our $t_0$ determination protocol is robust. The fluence independence of our results ensures that this static offset in excitation size does not influence the underlying physics extracted from the MSD. In all cases we see that the MSD, as plotted in Figure 2c, initially increases rapidly, with a maximum increase of 173 nm for the n = 6.5 microcavity and smallest increase (22 nm) for n = 3.5. Unsurprisingly, the degree of MSD expansion is correlated with the period of time over which the expansion occurs, from 270 fs for n = 6.5 down to 25 fs for n = 3.5, although we note this latter quantity approaches the limit of our instrument resolution.

*Coherent propagation model.* To convert this population expansion into meaningful velocities, we build a model of coherent polariton propagation and project the population density onto an approximately Gaussian profile. We note that our excitation conditions additionally excite intracavity exciton states, but they exhibit negligible propagation on these timescales and can be treated separately, see below. At $t_0$ due to the Gaussian excitation spot of the pump pulse, the polariton population is Gaussian in real space. As the photon dispersion is naturally inversion symmetric and the excitonic dispersion is essentially flat, the LP branch can be modelled as having an inversion symmetric one-dimensional energy-momentum dispersion relation about k = 0. The underlying Gaussian population induced by the pump is therefore composed of equal populations of polaritons from ±k with equal but opposite group velocities $\mp v_g$ (see **SI, S6**). Hence at early times, before scattering and diffusive behaviour dominate, we can decompose our initial excitation's time evolution into two Gaussian subpopulations that move at $\mp v_g$ and the overall dynamics can be modelled as,

$$n(x,t) = \frac{1}{2\sigma_0\sqrt{2\pi}}\left(\exp[-\frac{(x-v_g t)^2}{2\sigma_0^2}] + \exp[-\frac{(x+v_g t)^2}{2\sigma_0^2}]\right). \text{ (Equation 2)}$$

From equation 2 we can see that the overall polariton density profile will remain approximately Gaussian if the term $v_g t/\sigma_0$ is small, as is the case in the data in Figure 2. We therefore Taylor-expand Equation 2 in $vt/\sigma_0$ and find that the lowest non-trivial order is 2. We then reference against the 2$^{nd}$ order in $vt/\sigma_0$ truncated expansion of Equation 2 to a Gaussian density profile with a temporally varying $\sigma(t)$,

$$n(x,t) = A\exp[-\frac{(x-\mu)^2}{2\sigma(t)^2}], \text{ (Equation 3)}$$

as this is the experimentally relevant density profile and its form permits easy comparison to the diffusive regime, where $\sigma(t)^2 = \sigma_0^2 + 2Dt$ is expected. We then arrive at the result,

$$\sigma(v_g, t) = \sigma_0 + \frac{v_g^2 t^2}{\sigma_0}. \text{ (Equation 4)}$$

The initial coherent polariton population includes both UP and LP states with a range of k-vectors covering ±64.2°. We expect that this population and its associated transport to persist until it is lost to photon emission (chiefly governed by the cavity photon lifetime) or scattering to localised excitonic dark states (related to the exciton fraction of the wavefunction). To account for these effects, we incorporate a generalized coherence loss term into our expansion expression. We model this as an exponential decay with a combined scattering and cavity photon loss characteristic time constant $\tau$, yielding the result,

$$\sigma(v_g, t) = \sigma_0 + \exp(-t/\tau)\frac{v_g^2 t^2}{\sigma_0}. \text{ (Equation 5)}$$

This equation expresses coherent transport of polaritons, limited only by their decoherence into pure photonic or excitonic states. Fitting Equation 5 to the experimental data in Figure 2c, we find excellent agreement with the measured $\sigma(t)$ and are able to extract $v_g$ and $\tau$ from our fits for every cavity (Figure 2d,e). The fits provide scattering times increasing from ~20 fs to ~200 fs with increasing $Q$-factor, in accord with the expected behaviour for polaritons. The corresponding mean polariton velocities range from $0.12 \times 10^6$ m s$^{-1}$ to $0.65 \times 10^6$ m s$^{-1}$, (n =3.5, $v_g = 0.12 \pm 0.01 \times 10^6$ m s$^{-1}$; n = 4.5, $v_g = 0.36 \pm 0.1 \times 10^6$ m s$^{-1}$; n = 5.5, $v_g = 0.55 \pm 0.04 \times 10^6$ m s$^{-1}$; n = 6.5, $v_g = 0.65 \pm 0.2 \times 10^6$ m s$^{-1}$) orders of magnitude faster than purely excitonic transport in such systems.

For comparison we have calculated the expected group velocity for our microcavities. As discussed in the SI (S1), our broadband and full-angle excitation conditions result in simultaneous excitation of a wide range of states along the UP and LP branches. We approximate the population distribution initially generated through a combination of the pump pulse spectral profile and the angle-dependent photonic fraction of each branch extracted from a coupled-oscillator fit. We then apply this distribution to generate a weighted-average group velocity. We treat the UP and LP separately to represent upper and lower limits of the expected transport behaviour. The triangles in Figure 2e represent averages over 200-215 nm active-layer thicknesses, which represents the limit of the variation we expect based on our experimental procedure. We find that our observed transport velocity is in all cases markedly smaller than the predicted group velocity. Moreover, the short-lived phase of fast transport is followed by a collapse of the spatial profile back to the original $t_0$ dimensions. This behaviour is a sign that our measurement captures two distinct populations: coherent polaritons which decay through photonic leakage or scattering into the dark states, and highly localized dark exciton states which do not move on the timescale of the experiment. The latter include intracavity excitons directly excited by the pump pulse. Strikingly, our results reveal that within the initial coherent regime the polariton velocity systematically increases with cavity $Q$-factor. This behaviour is directly evident in the experimental traces and fully independent of the type of model applied to extract the velocities.

**Discussion**

This effect is challenging to rationalise. Indeed, the $Q$-factor *a priori* should have no effect on the strength of light-matter coupling: the identical active layer structure means we attain comparable Rabi splittings ~120 meV for all microcavities in our series, and the gradient of the LP dispersion which ultimately governs the polariton velocity varies little within the 200-215 nm active layer thickness range. The predicted group velocities in Figure 2e thus exhibit no meaningful $Q$-factor dependence; the slight upward trend in $v_g$(UP) is due to improved distinction of the TE-TM splitting at higher angles, which slightly biases the fitting. In the following we discuss potential explanations for our experimental observations, which to the best of our knowledge do not fall under any standard model of molecular polaritons.

The $Q$-factor should directly impact the polariton lifetime, which could provide increased opportunities for the excitation to scatter from the polariton branches into localized high-k molecular states. We accordingly expect an enhanced time-averaged LP-to-dark state scattering probability in our higher-Q structures. Once these localized states have been populated, subsequent transfer back to the LP through established scattering channels can harness thermal energy to generate higher-k (and potentially higher-velocity) polariton states.[60] If active, these channels should result in a greater steady-state spatial extent of polariton propagation as the *Q*-factor increases. However, the dynamics of these processes are limited by the slow scattering from dark states to the LP, typically of order 1-300 ns$^{-1}$,[22,37,45] and thus the established pathways of molecular polariton photophysics cannot explain our observations of ultrafast, *Q*-factor-dependent transport.

***Higher Q causes reduced disorder and enhanced delocalization.*** Instead, as we lay out below, we propose that an increased cavity *Q*-factor reduces the energetic disorder and increases the delocalization of the intracavity dark states, and that these play a substantial role in polariton transport even on early timescales. The principal effect of increasing *Q*-factor is to narrow the bandwidth of the photonic density of states. In the case where the photonic bandwidth is narrower than the excitonic bandwidth—a typical condition for organic exciton-polaritons—only a subset of the absorbers represented by the inhomogeneously broadened exciton absorption will enter the strong-coupling regime. Specifically, it is the absorbers most resonant with the photonic mode which are most likely to strongly couple, resulting in a distribution of polaritons with reduced linewidth compared to the parent excitons. The N-1 dark states that correspond to the bright polaritons formed must likewise exhibit significantly reduced energetic bandwidth relative to the uncoupled states. It directly follows that our higher-*Q* cavities couple to progressively energetically narrower ensembles of molecular states. Crucially, this means that there is significantly lower energetic disorder within the manifold of dark states.

Building on these ideas, we propose a correlation between the 'disorder amplitude' (set here by the energetic bandwidth of the states and the *Q*-factor) and the localisation length, which parameterizes how delocalised the states in the system are and how far/fast transport occurs. We emphasise that this 'disorder amplitude' is distinct from the classical inhomogeneous disorder of traps *etc*, which was previously used to explain the lower than expected velocity of exciton-polaritons in all-metal J-aggregate cavities,[32] and is purely related to the bandwidth of the dark states. We suggest that reduced energetic disorder within such dark states suppresses the processes of decoherence or localization onto single molecular sites as the coupling between the dark states is enhanced.

In both pure semiconductor and pure photonic systems, disorder and delocalisation have been shown to be intricately linked.[61] We extend this concept to propose the dark states are no longer fully localized. Indeed, while the dark states are typically treated in terms of localized molecular states,[35] they can equivalently be described as collective excitations across the ensemble, albeit with no photonic component.[34,62] In certain cases, these dark states are theoretically predicted to achieve comparable spatial delocalization to the bright polaritons and can be involved in efficiently transferring excitations.[63] Other theoretical work reveals that the delocalization of dark states is directly enhanced by narrowing the photonic density of states, resulting in an efficient contribution to the coherent transport of bright polaritons.[64] The latter proposal is remarkably similar to the experimental observations seen here.

*Rapid dark state-to-LP exchange.* Because of their spatial extent, these states should have relatively favourable wavefunction overlap with the bright polaritons, increasing the rate of population exchange between the two manifolds. This rapid population transfer is the essential component of our model, necessary to link the property of ballistic transport (characteristic of polaritons) with the *Q*-factor dependence (which affects the dark states) within our sub-ps observation window. Indeed, if we assume the dark states are stationary, our results require strikingly high rates of dark-to-polariton population transfer of >50 ps$^{-1}$. This value is several orders of magnitude higher than the rates typically invoked.[34,37,45] However, it has been suggested that energetic proximity can enable the UP, LP and dark states to persist in rapid equilibrium for well beyond the intrinsic polariton lifetime,[65] and the linewidth of BODIPY-R should make that regime relevant here. In our framework, the coherent dark states function simultaneously as a brake—reducing the transport velocity for the proportion of its lifetime that the population spends in the dark manifold—and as an efficient reservoir to regenerate the highly propagating polariton states. The system then exhibits polariton-enhanced transport until decoherence sets in with a characteristic lifetime that presumably contributes to our phenomenological $\tau$, beyond which point only localized molecular states are populated.

While the model can fully describe our unusual observations, the estimated rates of population exchange pose a further challenge. Typical values for transfer from dark states to the polaritons are on the order of 1-300 ns$^{-1}$ if not slower.[22,34,44,45,66] The critical distinction is that these values arise from radiative pumping and vibrational scattering mechanisms in which the initial state is incoherent and localized on a single molecule.[12,37,38] We suggest that a different mechanism is at play on early timescales, when the dark states are still coherently delocalized. Similar rapid equilibration between polaritons and dark states has been proposed based on multiscale molecular dynamics calculations.[65] Further support for this idea comes from studies of polariton-mediated donor-acceptor energy transfer and ultrafast polariton relaxation dynamics. The intracavity dark states are widely expected to mediate energy transfer between strongly coupled donors and acceptors,[17,28] yet the timescales of this transfer are substantially faster than anticipated from the standard incoherent mechanisms.[26,27,67] In single-component systems, high time-resolution two-dimensional electronic spectroscopy suggests a <100 fs component of dark-to-polariton population transfer.[54]

Our qualitative model additionally provides a route to rationalize comparison with an earlier TAM study of organic exciton-polariton transport. The velocities we capture in the early coherent regime closely match those reported for very low-*Q* TDBC J-aggregate microcavities on the 1-10 ps timescale.[32] However, according to standard models, on those timescales the system should have been firmly within the incoherent, reservoir-dominated regime where transport is dominated by local hopping. We consider that the key difference between these systems is the form of the active layer: well-dispersed BODIPY-R molecules in the present work, versus J-aggregates boasting highly delocalized, coherent exciton states in the earlier study. Within our model, these intrinsically coherent exciton states within the J-aggregates serve a comparable role to the delocalized dark states in our system, enabling effective interaction with the rapidly transporting polaritons due to favourable wavefunction overlap. Their exciton properties mean they continue this rapid dark state-polariton exchange well beyond typical localization timescales (<<50 fs in the reported structures). This suggests two distinct routes to enhance the transport properties of exciton-polaritons: exploit more highly delocalized exciton states within the active layer, or enhance the cavity $Q$-factor to impart delocalization.

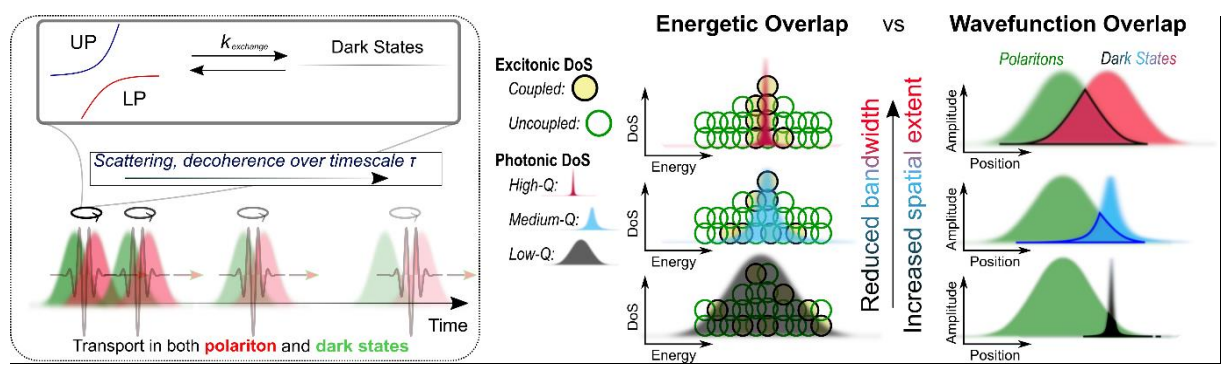

**Figure 3: Summary of proposed mechanism for *Q*-factor dependence of transport.** (Left) Rapid population exchange between bright polaritons and dark states occurs at a rate $k_{\text{exchange}}$. Until decoherence, these states transport together. (Centre) The $Q$-factor directly impacts the photonic density of states (Gaussians) and thus indirectly impacts the energetic distribution of coupled vs uncoupled molecules (circles), though the same proportion of molecules are strongly coupled in every case. The resulting dark states exhibit $Q$-factor-dependent bandwidth. (Right) This effect leads to correspondingly $Q$-factor-dependent delocalization of the dark states (black, blue, red). As the spatial extent of the bright polaritons (green) is large and approximately constant, this alteration of dark states is the principal means to alter the wavefunction overlap (outlines) and thus rate of exchange between polaritons and dark states.

## Conclusion

Using fs-TAM, we have directly tracked the spatial propagation of organic exciton-polaritons within their initial coherent lifetime. This sub-ps regime is infrequently characterized but essential to distinguish the intrinsic properties of polaritons from their complex interactions with incoherent dark states. Our results provide evidence of two distinct excited populations: rapidly propagating polaritons and a reservoir of localized intracavity dark states that scarcely move within the experimental timescale. Using a phenomenological model of band-like transport of an initially Gaussian distribution, we obtain velocities in the range of $0.1 \times 10^6$ m s$^{-1}$ to $0.65 \times 10^6$ m s$^{-1}$. Through variation of the cavity structure, we determined that the polariton transport velocity is directly enhanced with increasing cavity $Q$-factor.

This behaviour is not captured by current models of exciton-polaritons. Aside from the straightforward effect on the polariton (and thus transport) lifetime, the $Q$-factor has negligible direct effects on the polariton states. Instead, current theoretical work suggests that only the intracavity dark states should be meaningfully affected through modification of their energetic bandwidth and localization properties.[64] We thus reconcile our observed dependence and the unexpectedly low polariton velocities in terms of rapid population exchange/overlap between these two species. The dark states do not necessarily 'move' themselves, but nor are they localised at early times as classically held. Instead, the dark states are delocalised by an amount set by the cavity $Q$-factor. This delocalisation and overlap with the bright polariton states imparts a $Q$-factor dependence on the transport velocity and reduces the polariton velocity from that expected based on the dispersion alone.

Further work is required to disentangle the impact of material disorder on the properties of bright and dark states within such microcavities and reveal the mechanistic details of the interactions we propose. Nonetheless, our results highlight that intracavity dark states can play a substantial role even in the ultrafast dynamics of processes that are intrinsically polaritonic, such as long-range transport. Crucially, this role changes dramatically over the course of scattering and localization, demonstrating the importance of studying molecular polaritons within their coherent lifetime. The signatures of the dark states have been persistently challenging to identify spectroscopically in exciton-polariton systems,[26,27,46,48,51,68] despite their central role in polariton physics and significantly higher population.[34] Indeed, the transient absorption spectra in Figure 1 exhibit only subtle changes between coherent and reservoir-dominated regimes. However, the stark difference in transport properties on early and long timescales provides a unique handle to interrogate their properties. This approach opens new avenues to explore how these essential states may be tuned and their contribution to polaritonic physics controlled, as well as highlighting the ability to tune disorder and transport in room-temperature molecular polaritonic systems.

**Methods**

Sample preparation

The organic dye BODIPY-R was synthesized following published protocols.[69] For thin film and microcavity preparation, we dissolved the dye to a concentration of 2.5 mg/mL in toluene. To increase processability and minimize aggregation and quenching effects, the dye was co-dissolved with polystyrene (PS, molecular weight 192000, Sigma Aldrich) at a concentration of 25 mg/mL, corresponding to ~10% dye in PS by weight. This solution was spin-coated onto 0.17 mm thin coverslips or the bottom mirrors of Fabry-Perot microcavities (assembled on 0.17 mm thin cover glass slides). To prepare the dielectric microcavities $SiO_2$ and $Nb_2O_5$ were deposited with e-beam evaporation, with layer thicknesses of 105 nm ($SiO_2$) and 71 nm ($Nb_2O_5$) in the structure

Nb$_2$O$_5$/(SiO$_2$/TiO$_2$)$_5$ to yield distributed Bragg reflectors (DBRs) with a stop-band centred at 635 nm. Top and bottom DBRs were fabricated using identical evaporation parameters. In both structures, top mirrors were deposited directly onto the spin-coated organic layer, which was tuned in the range ~215-240 nm for DBR structures to yield λ/2 cavities.

Absorption, emission and reflectivity

UV-Vis absorption of the BODIPY-R control films was performed on a Horiba Fluoromax 4 fluorometer utilised with a Xenon lamp. Control films were excited using a CW 473 nm laser diode with emission imaged into an Andor Shamrock SR-193i-A double grating imaging spectrograph. Angle-resolved white reflectivity measurements were carried out using a goniometer setup equipped with two rotating arms. A halogen–deuterium white light source (DH-2000-BAL) was focused on the sample's surface using optics mounted on the excitation arm. Reflected light was passed through a series of optics mounted on the collection arm and sent into an Andor Shamrock SR-303i-A CCD spectrometer using a fibre bundle.

Femtosecond pump-probe spectroscopy

The fs-TA experiments were performed using a Yb-based amplified system (PHAROS, Light Conversion) providing 14.5 W at 1030 nm and 38 kHz repetition rate. The probe beam was generated by focusing a portion of the fundamental in a 4 mm YAG substrate and spanned from 520 nm to 1400 nm. The pump pulses were generated in home-built noncollinear optical parametric amplifiers (NOPAs), as previously outlined by Liebel *et al*. The NOPA output (~4 to 5 mW) was centred typically between 520 and 560 nm (FWHM ~65-80 nm), and pulses were compressed using a chirped mirror and wedge prism (Layerterc) combination to a temporal duration of ~9 fs. Compression was determined by second-harmonic generation frequency-resolved optical gating (SHG-FROG; upper limit) and further confirmed by reference measurements on acetonitrile where the 2200 cm$^{-1}$ mode could be resolved. The probe white light was delayed using a computer-controlled piezoelectric translation stage (Physik Instrumente), and a sequence of probe pulses with and without pump was generated using a chopper wheel (Thorlabs) on the pump beam. The pump irradiance was set to a maximum of 38 μJ/cm$^2$. After the sample, the probe pulse was split with a 950 nm dichroic mirror (Thorlabs). The visible part (520–950 nm) was then imaged with a Silicon photodiode array camera (Entwicklunsbüro Stresing; visible monochromator 550 nm blazed grating). The near infrared part was imaged using an InGaAs photodiode array camera (Sensors Unlimited; 1200 nm blazed grating). Measurements were carried out with a time step size of 4 fs out to 2 ps to minimize the exposure time of the sample to the beam. Unless otherwise stated, all measurements were carried out with the probe polarisation set parallel with respect to that of the pump (using a half-waveplate; Eksma). The absorption spectrum of samples was measured after each pump-probe sweep to account for any sample degradation.

Femtosecond pump-probe microscopy

The femtosecond wide-field detected transient absorption microscope has been described in detail previously. Briefly, a Yb:KGW amplifier system (LightConversion, Pharos, 5 W, 180 fs, 1030 nm, 200 kHz) was used to seed two white-light stages for pump and probe generation. The pump white-light (3 mm Sapphire) was spectrally adjusted with a 630 nm short-pass tunable 'razor-edge' filter (Semrock) and compressed to 10 fs for all optical elements with two pairs of third-order compensated chirped mirrors and a wedge-prism pair (Layertec). Subsequently, the mode of the pump pulse is cleaned by a pinhole before being focused through the objective lens (NA = 1.1, oil immersion) to a spot size of ~260 nm (full-width-half-maximum). The probe white-light (3 mm YAG) was spectrally adjusted to 635 – 900 nm in a home-build fused silica prism filter and compressed to 7 fs with a pair of third-order compensated chirped mirrors and a wedge-prism pair (Venteon) before being free-space focused onto the sample (20 micron Gaussian spot size full-width-half-maximum). The transmitted probe was imaged onto an emCCD (Rolera Thunder, Photometrics) at 55.5 nm/pixel as verified by a resolution target. The frame rate of the camera was set to 30 Hz with an integration time of 11 ms and pump off/on images were generated by a mechanical chopper at a frequency of 15 Hz.

**Supporting Information**

Optical modelling, transient absorption dynamics, transient microscopy measurements on control samples, characterization of the excitation spot size, and details of the coherent transport model.

**Acknowledgements**

The authors thank David G. Lidzey for useful discussion. We acknowledge financial support from the EPSRC and Winton Program for the Physics of Sustainability. R.P. additionally thanks the EPSRC for a Doctoral Prize Fellowship and Nicolas Gauriot (Cambridge) for assistance with fs-TA measurements. A.J.M. and S.J.R. were supported by the U.S. Department of Energy, Office of Science, Basic Energy Sciences, CPIMS Program under Early Career Research Program award number DE-SC0021941.

**Associated Content**

The data associated with this manuscript is freely available at [url to be added in proof].